\documentclass{llncs}
\usepackage{makeidx}  
\usepackage{graphicx}
\begin{document}

\spnewtheorem*{<env_nam>}{<caption>}{<cap_font>}{<body_font>}

%
%
\pagestyle{headings}  
\addtocmark{Hamiltonian Mechanics} 

\mainmatter              
\title{Is Evaluating Visual Search Interfaces in Digital Libraries Still an Issue?\thanks{\textit{Copyright \copyright{} $2014$ by the paper's authors. Copying permitted only for private and academic purposes.} In: T. Seidl, M. Hassani, C. Beecks (Eds.): Proceedings of the LWA 2014 Workshops: KDML, IR, FGWM, Aachen, Germany, 8-10 September 2014, published at http://ceur-ws.org}}
\titlerunning{Hamiltonian Mechanics}  
%
\author{Wilko van Hoek \and Philipp Mayr}
%
%
%
\institute{GESIS - Leibniz Institute for the Social Sciences, Unter Sachsenhausen 6-8, 50667 Cologne, Germany\\
\email{\{wilko.vanhoek,philipp.mayr\}@gesis.org},\\ WWW home page:
\texttt{http://www.gesis.org}}

\maketitle              

\thispagestyle{empty}
\pagestyle{empty}

\begin{abstract}
Although various visual interfaces for digital libraries have been developed in prototypical systems, very few of these visual approaches have been integrated into today's digital libraries. In this position paper we argue that this is most likely due to the fact that the evaluation results of most visual systems lack comparability. There is no fix standard on how to evaluate visual interactive user interfaces. Therefore it is not possible to identify which approach is more suitable for a certain context. We feel that the comparability of evaluation results could be improved by building a common evaluation setup consisting of a reference system, based on a standardized corpus with fixed tasks and a panel for possible participants.
\end{abstract}

\section{Introduction}

In the last twenty years of research on visual interfaces for digital libraries (DLs) a variety of approaches has been proposed and many visual search prototypes have been developed to support the user of DLs in his search process. For every part of the search process techniques exist to support the user. However, most of these techniques have not found their way into today's DLs. On the contrary nearly all prototypes have been discontinued. Most ideas have not been evaluated more than once in a relatively small study.

The main question is: Why have most of the research results not been adapted into today's DLs? One simple answer could be that this is the typical evolution of scientific research. Many ideas are not supposed to be commercially beneficial, adaptable in large scale live environments or not successful due to various other reasons. In \cite{vanHoek_assessing_2014} we took a look at the different techniques and the studies that have been conducted, so far we do not feel that the answer is that simple.

After taking a closer look at the results of a line-up of different studies \cite{nowell_visualizing_1996},\cite{sebrechts_visualization_1999},\cite{shneiderman_visualizing_2000},\cite{andrews_infosky_2002},\cite{granitzer_evaluating_2004},\cite{kim_exploring_2011},\cite{dork_pivotpaths_2012} and \cite{kodagoda_using_2013}, we can observe that usually quantitative results on the task performance and the accuracy of participants in a visual IR system are comparably poor or at least equally good as a strictly text-based system. On the other side, in accompanying questionnaires, the participants' opinions on the same visual IR system were positive and in favour of the system. Here seems to be a mismatch that needs a closer examination.

\section{State of the Art}

In the following we briefly review a selection of well-known publications which describe and evaluate information visualization systems for digital libraries \cite{vanHoek_assessing_2014}. The section will introduce seven prototype systems that provide a visual access to data and studies that were conducted to evaluate these prototypes. We will take a closer look at five different facets of how the studies were conducted. We will try to identify:
\begin{enumerate}
  \item the main aim of the study (measurement of usability, performance or the cognitive effects),
  \item the type of evaluation method that was used (e.g. A/B testing, between- or within-subject design),
  \item how the study was conducted (e.g. task-based, laboratory),
  \item details on the subjects (e.g. group size, expertise),
  \item the document corpus that was used (e.g. newspaper articles, digital libraries). 
\end{enumerate}

\subsection{ENVISION}

The ENVISION system \cite{nowell_visualizing_1996} is an early attempt to display search results in a 2-dimensional grid. Metadata fields like author or publication year could be selected for the two axes and the system would position the search results represented by icons within the resulting grid. 

In the study that was conducted, the main aim was to evaluate the usability of such a system. This was done without A/B testing. The users were asked to fulfill several tasks that involved using different interaction methods in the system. The tasks were not aligned with those of other studies. The study took place in a laboratory environment with one expert, two graduate students and two undergraduate students. As document corpus, scientific publications were used. There are no further details about the corpus.

\subsection{NIRVE}

\begin{figure}[ht!]
\centering
\includegraphics[width=\linewidth]{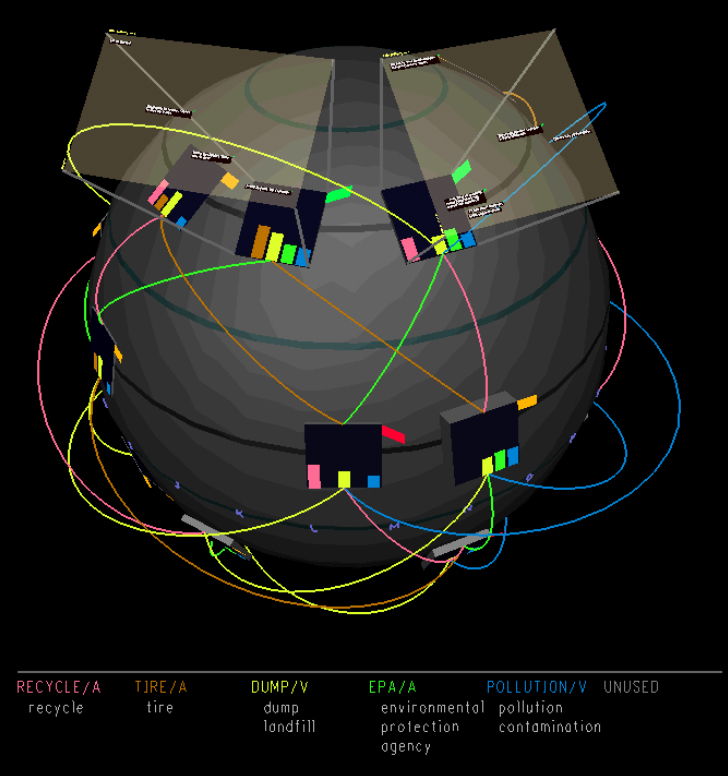}
\caption{Globe view of the NIRVE sytem}
\label{fig:nirve}
\end{figure}

In the NIRVE system \cite{sebrechts_visualization_1999}, search results can be displayed on a 3-dimensional globe, where clusters of documents are displayed as boxes emerging from the globe. The thickness of a box represents the number of documents in the cluster. Documents with the same combination of query terms build a document cluster. Clusters of documents containing only a few query terms are displayed near the south pole, clusters of documents containing more query terms near the north pole (cf. figure \ref{fig:nirve}).

In the study that was conducted to asses the system, a text-based IR system a 2-dimensional version of the globe and the globe-based system were compared. The aim was to assess usability and performance of the globe system. The performance was measured between the three systems. The study was conducted in a task-based laboratory environment. The tasks were not aligned with those of other studies. The subject group consisted of 15 participants of which 6 were experts and 9 students. The underlying corpus consisted of the news stories of the Associated Press from the year 1988.

\subsection{GRIDL}

In the GRIDL system \cite{shneiderman_visualizing_2000} the search result list is presented in a 2-dimensional grid similar to the presentation in ENVISION \cite{nowell_visualizing_1996}. Here the focus lays on overcoming the problem of overcrowded rows, columns or cells. An attempt was made to overcome this issue by utilizing solutions such as tool tips or further hierarchical grouping.

Two consecutive studies were conducted on the system. The main aim was to assess the usability of the system. The studies were done in a task-based laboratory environment without A/B testing. The tasks were not aligned with other studies. The first subject group consisted of 8 graduate students and the second of 24 subjects, of which 10 came from the field of library science, 8 from the field of computer science and 6 subjects from other fields. As corpus, metadata of scientific publications within the database of the Computer Science Department Library at the University of Maryland was used.

\subsection{InfoSky}

The InfoSky system \cite{andrews_infosky_2002} provides the user with two different alternatives to browse and query large data sets of hierarchically structured documents. The first one is a tree browser similar to the file browser in operation systems. The second one is a so-called telescope browser. In the telescope browser, documents are represented as points on a black background modelled after the night sky. The documents hierarchy and cosine similarity are 
used to position and group the documents. In this way clusters are formed, consisting of documents that bear a certain resemblance to each other.

The system was evaluated in a first study \cite{andrews_infosky_2002}. Based on the results of this first study the system has been improved and extended. It then was evaluated in a second study \cite{granitzer_evaluating_2004}. The aim of both studies was to assess the performance of users using the telescope browser. Therefore A/B testing with a crossover design was used to assess the user performance with telescope and tree browser. Both studies were conducted in a task-based laboratory environment. The tasks were not aligned with other studies. Moreover, the tasks were changed for the second study. The first study took place with 8 subjects, the second with 9 subjects. No further details on the background of the participants were provided. The corpus of both studies was a set of 80,000 newspaper articles from the German Sueddeutsche Zeitung.

\subsection{VIDLS}

\begin{figure}[ht!]
\centering
\includegraphics[width=\linewidth]{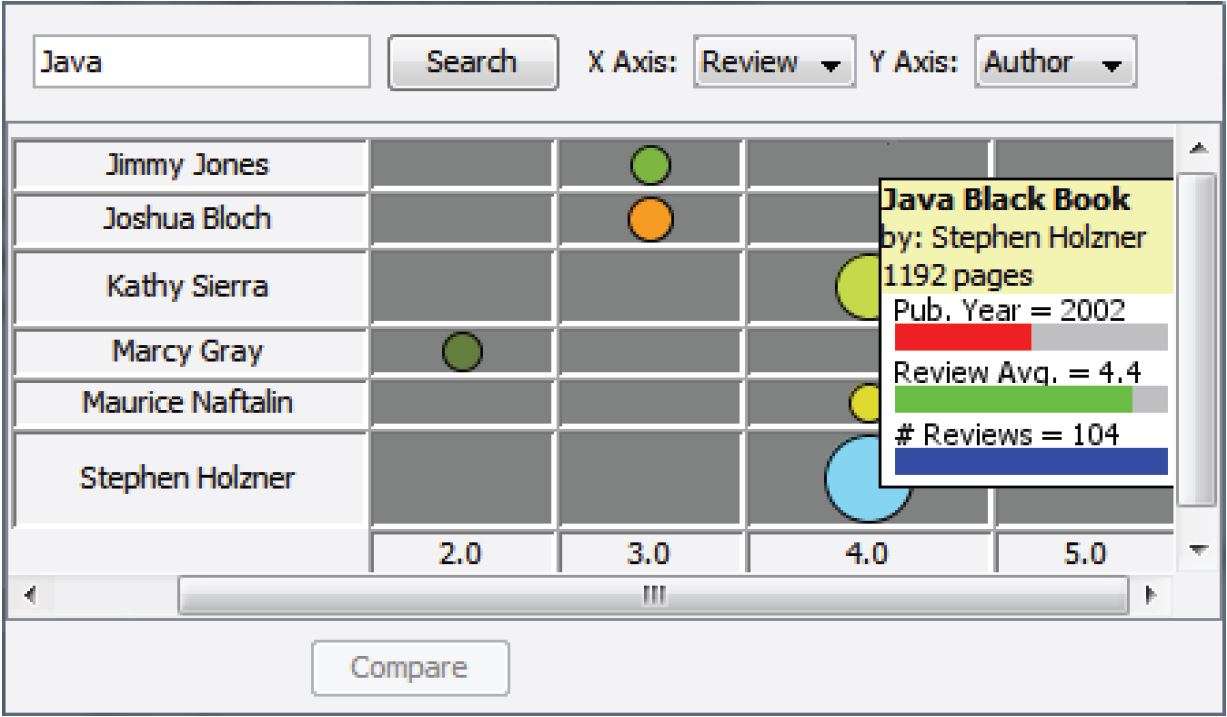}
\caption{Result list overview in the VIDLS system}
\label{fig:vidls}
\end{figure}

In the VIDLS System \cite{kim_exploring_2011}, three visual interfaces have been implemented. An overview for the result list of searches, and two detailed document views. The system relies on full text documents of books, as the visualization uses the content and the index of the documents. The overview uses a 2-dimensional grid layout following the GRIDL System \cite{shneiderman_visualizing_2000}. Here books are represented as circles. The size of the circle resembles the normalized number of pages on which the search terms occur (cf. figure~\ref{fig:vidls}). In the detailed view, the book's index is used to display distribution and frequency of index terms and search terms within the document.

The main aim of the study was to assess the usability of the visualizations. Therefore A/B testing with a within-subject design was used. This was in a laboratory environment. The study was divided in multiple sessions each with three to five students. The only task was to search for books, once with both systems, the text-based system and the VIDLS system. A post search survey was used to assess the usability by asking the users about their impressions on the system. No details on the corpus are provided.

\subsection{PivotPaths}

In the PivotPath system \cite{dork_pivotpaths_2012} the search result list as visualization canvas is presented as an information space that contains multiple facets and relations, such as authors, keywords, and citations of academic publications, or actors and genres of movies. PivotPath focuses on selecting items from facet lists (pivot operations) resulting in direct changes on the interface. The PivotPaths interface exposes faceted relations as visual paths in arrangements that invite the viewer to 'take a stroll'.

Two participants' observational studies (academic publications and movie collections) were conducted in an intranet deployment of the system. The authors did semi-structured interviews where participants could comment on questions and executed tasks. The studies were done in a task-based laboratory environment without A/B testing. The tasks of the user sessions were not aligned with other studies. The intranet study attracted 290 participants with 211 actual-use sessions. The authors report detailed on anecdotal email feedback of their participants. As corpus, metadata of scientific computer science publications from Microsoft Academic Search and movies from the Internet Movie Database were used.

\subsection{INVISQUE}

In the INVISQUE system \cite{kodagoda_using_2013} the query formulation and result list presentation has been moved into one interface. In this system search results of different search can be displayed on an infinite pane. The result sets of different searches can be merged by dragging one onto another. In this way complex boolean queries can be generated on a visual level by working with the result sets of queries.

In the study the decision was made not to evaluate performance or usability, but to assess the sense-making process of experts using the system. Therefore six senior university librarians were asked to identify three central authors of a field that was unknown to them. Interaction-logs, video recording and survey were analyzed to evaluate the study. As corpus the metadata of publication from the ACM SIGCHI conference from 1982 to 2011 was used.

In the following section we will develop and discuss positions which we think are still crucial in the domain of visual search interface in DL. We will emphasize to develop a more standardized evaluation setup for such interfaces. We are aware of the fact that an experimental approach has already been implemented during the TREC interactive tracks (TREC 3-12) \cite{voorhees_trec_2005} that follows the some identical arguments and observations we are discussing in this paper and thus, that parts of the following positions have already been discussed. Especially in TREC-6 an almost identical approach has been applied to assess cross-site performance \cite{lagergren_comparing_1998}. The results of this analysis were mixed. It was not possible to reliably compare the performances of the different systems. It was emphasized that by further investigating cross-sites experiments more reliable methods could be generated. Also, the most problematic factors influencing the results of the comparison were the relevance assessor’s and the fact that the subjects differed throughout the different studies. We therefore think that the ideas and findings that have led to developing a cross-site analysis are still relevant and in our analysis we could still identify those shortcomings in interface evaluation methodology even in more recent studies. 

\section{Discussion}

\textbf{Position A: Diversity of evaluation aims.} Throughout all studies we could see that there has been a clear aim that was followed. Usability and performance are two central aspects of systems, but as \cite{dork_pivotpaths_2012} and \cite{kodagoda_using_2013} have shown, there are other aspects that are important when it comes to the question of the suitability of an interface for a DL. Anyhow, except for \cite{dork_pivotpaths_2012} and \cite{kodagoda_using_2013}, we are missing a real discussion about why usability or performance was considered to be more relevant than the other aspects. Also \cite{dork_pivotpaths_2012} and \cite{kodagoda_using_2013} make clear that they are interested in other aspects, but then they ignore usability and performance completely. A system that is hard to use cannot be considered to optimize performance or serendipity. In addition, performance has its influence on other aspects as well. We strongly feel that the various aspects of the systems are co-dependent on each other. Instead of assessing only one aspect, multiple aspects should be assessed. At least a usability and a performance study should be conducted. We do see that this implies a more complex study design and costs more effort. However, this might be compensated by creating a standardized evaluation design and environment.

\textbf{Position B: Missing shared design methodology.} When thinking of standardization of the evaluation, one needs to decide which study design to use for which types of study. In all cases where performance was measured, A/B testing was used. Obviously this is a good idea, as performance implies benchmarking, which does not make a lot of sense without reference values. These reference values can be generated by measuring the task performance in a reference system. Usability on the other side can be assessed without a reference system. There might be a way to include a usability study into a performance study and to reduce the need of conducting two separate studies. In our review we observed a missing shared design methodology which would be very essential to reach comparability and reproducibility in this domain.

\textbf{Position C: Need of a common reference system.} In total A/B testing seems to be an important tool to evaluate system. But are results of A/B testing really worth the effort of comparing two systems? In an ideal world, one would assume that when comparing systems A and B and systems A and C one could make assumptions about the relation between B and C. But when A is not fix this transitivity is lost. We have seen A/B testing being used in \cite{sebrechts_visualization_1999},\cite{andrews_infosky_2002}, and \cite{kim_exploring_2011}. In all three studies an own implementation of a text-based system was used as reference. There is no clarification in how far the three text-based systems are comparable. Thus, we do not know anything about the relation between the three prototypes. We do not know which changes have improved or worsened the usability or performance of a grid-based visualization in comparison to a text-based system. We therefore propose to build a common reference system. This would be a more suitable baseline for evaluating system in an A/B testing scenario. During the TREC-6 evaluation it was impossible to make the reference system and the different experimental system accessible from the same spot. Therefore the participants needed to implement their reference system at their own institute to conduct the user studies \cite{lagergren_comparing_1998}. With today’s digital infrastructure it would be easier to access a reference system remotely. Thus the effort of conducting a study with a reference system is reduced significantly and it would be possible to test two experimental systems A and C and a reference system B in the same study with the same subjects.

\textbf{Position D:  Need for standardized test collections.} If we follow the trail of thought in our position C, it becomes clear that building a reference system is not enough to improve comparability. The results of a study are also influenced by other factors. A visualization technique might be suitable for a certain set of documents, but not even applicable for another. Thus, the reference system should be combinable with different document corpora. But using different corpora for similar systems is not a good idea. In \cite{sebrechts_visualization_1999} and \cite{andrews_infosky_2002} for example, the underlying corpus was a set of newspaper articles, but not the same set.  This degrades the comparability of the study results. In addition, in TREC-6 of the TREC interarctive tracks only one collection (The Financial Times of London 1991-1994 collection) was used. This collection was not suitable for all experimental interfaces as they focused on different aspects.  We feel it might be a good idea to create a set of standardized test collections, so that similar systems can be assessed in the same environment. 

\textbf{Position E: Need of shared and standardized tasks.}
Another crucial point regarding the comparability of task-based studies are the tasks themselves. As long as every study defines its own tasks, it is not possible to compare the results easily. Aligning tasks is not a trivial task as different systems aim at different steps of the search process. On the other hand, the systems ENVISION \cite{nowell_visualizing_1996}, GRIDL\cite{shneiderman_visualizing_2000}, and VIDLS \cite{shneiderman_visualizing_2000} for example, all display search results in a 2-dimensional grid and refer to each other. Here arises the question, why are there no tasks that were aligned with previous studies? Building up a set of tasks for typical activities in DLs, so that researchers can compare the usability and performance of different systems is a next desideratum.

\textbf{Position F: Subject-based evaluation.}
The last issue we would like to address is the question on the subject group. In a laboratory environment, it is expensive to have many subjects. Experts are more difficult to get for a study than students. Throughout the eight studies we have seen a variety of expertises and sizes of the subject groups. In how far can results be comparable when the subject groups vary that much? What can be done to improve this situation? One way could be to establish a panel of subjects, that allows to contact the same set subjects for multiple studies (see e.g. \cite{Kern2014}). This seems to be a very important aspect. In \cite{lagergren_comparing_1998} it is argued that the cross-site analysis results are strongly biased by the differences in the subject groups which were involved at the various studies.

\section{Outlook}

In this paper we discussed some issues concerning the evaluation of visual interfaces for DLs. The main issue here is the comparability of results. We reviewed a set of studies on well-known interfaces. Comparing the studies we could identify some possible points for improvement. We propose to build up a common reference system, where methods and designs are fixed, based on standardized document corpora. The system should be evaluated with standardized corpora like the iSearch test collection \cite{lykke_developing_2010} or openly accessible publications from repositories like PubMed Central \footnote{http://www.ncbi.nlm.nih.gov/pmc/} or arXiv \footnote{http://www.arxiv.org}. In addition a set of tasks should be defined that reflect common information needs in DLs. Combined with a panel of participants \cite{Kern2014} a suitable environment to conduct more comparable studies could be created. Altogether, the development of such an environment is a complex and time-consuming project. This effort should be worked on collaboratively to benefit from the experiences of the researchers in the field. A workshop on a topic related conference like the IIiX or CHI is in preparation. The project could also benefit from collaboration with the TREC and CLEF groups to make use of their experiences with standardizing corpora and evaluation settings.

\section{Acknowledgement}

Part of this work has been funded by the COST Action TD1210 KnowEscape! http://knowescape.org/

\bibliographystyle{splncs03}
\bibliography{library}

\end{document}